\documentclass[12pt]{article}
\pdfoutput=1
\usepackage{subfigure}
\usepackage{amssymb,amsmath}
\usepackage{graphicx}
\usepackage{color}
\usepackage[colorlinks=true
,urlcolor=blue
,citecolor=blue
,linkcolor=blue
,pagecolor=blue
,linktocpage=true
,pdfproducer=medialab
]{hyperref}
\usepackage[a4paper,width=17cm]{geometry}
\makeatletter \renewcommand{\@dotsep}{10000} \makeatother
\usepackage{appendix}

\begin{document}

\begin{center}

 {\Large\bf  A Fundamental Form of the Schrodinger  Equation 
 } \vspace{1cm}

{   Muhammad Adeel Ajaib\footnote{ E-mail: adeel@udel.edu}}

{\baselineskip 20pt \it
Department of Physics and Astronomy, Ursinus College, Collegeville, PA 19426\\
 } \vspace{.5cm}

{\baselineskip 20pt \it
   } \vspace{.5cm}

\vspace{1.5cm}
\end{center}

\begin{abstract}

We propose a first order equation from which the Schrodinger equation can be derived. Matrices that obey certain properties are introduced for this purpose. We start by constructing the solutions of this equation in 1D and solve the problem of electron scattering from a step potential. We show that the sum of the spin up and down, reflection and transmission coefficients, is equal to the quantum mechanical results for this problem. Furthermore, we present a 3D version of the equation which can be used to derive the Schrodinger equation in 3D.

\end{abstract}

\newpage

\section{Introduction}\label{intro}

The problem of particle scattering from, for instance, a potential barrier or step are important fundamental examples in quantum mechanics. The Schrodinger equation can be employed to solve these problems in the non-relativistic limit. The Schrodinger equation, however, does not consider the spin of the particle in such scattering problems. 
For relativistic particles, the Dirac equation \cite{Dirac1} when applied to such problems leads to the Klein paradox \cite{Klein}. This paradox arises when the Dirac equation is used to study problems such as the scattering of an electron from a potential step, particularly when the barrier height $V_0 >2mc^2$. For this problem the reflected current turns out to be larger than the incident current. This paradox can be resolved by assuming the production of particle-antiparticle pairs at the barrier. Other methods have also been proposed in literature to resolve this paradox (see, for example, the list of references \cite{Hansen})

In this paper we introduce a non-relativistic first order equation which can be employed to include the spin of particles in scattering problems.  We first show that the Schrodinger equation can be derived from this fundamental first order equation in 1D. We derive the probability and current densities for this equation. Next, we construct the solutions to these equations and calculate the reflection and transmission coefficients for this process. We show that combining the results for spin up and down electrons leads to the quantum mechanical result for the reflection and transmission coefficients. Our analyses makes specific predictions about the manner in which spin up electrons can scatter off a step potential.

The paper is organized as follows: In section \ref{equation} we present a  fundamental form of the Schrodinger equation in 1D and derive the probability and current densities. We further discuss the plane wave solutions to this equation. With the solutions constructed, we solve the potential step problem for $E>V_0$ and $E<V_0$ in 1D in section \ref{step}. We introduce the 3D version of this equation and derive the probability current in section \ref{3d}.  Our conclusions are presented in section \ref{conclude}.


\section{An Equation Underlying The Schrodinger \\ Equation in 1D}\label{equation}

 In this section we present an equation that, even in the non-relativistic limit, can be seen as an underlying equation to the Schrodinger equation. The Schrodinger equation can be derived from this equation similar to the manner in which the Klein Gordon equation can be obtained from the Dirac equation. We restrict ourselves to a single spatial dimension (say, $z$) in this and the following sections. In section \ref{3d} we will present a 3D version of the equation. 
 
 We propose the following first order equation
\begin{eqnarray}
-i \partial_z \psi = (i  \eta \partial_t  + \eta^\dagger m) \psi
\label{mse-eq}
\end{eqnarray}
Here $\eta$ and $\eta^\dagger$ are matrices that obey certain relations described below. Using iteration the above equation leads to the following 
\begin{eqnarray}
(-i \partial_z )^2  \psi = (i \eta \partial_t)^2 \psi + i m \left\lbrace \eta, \eta^\dagger \right\rbrace  \partial_t \psi  + (\eta^\dagger)^2 m^2 \psi
\end{eqnarray}
In order to obtain the Schrodinger equation the matrices $\eta$ and $\eta^\dagger$ should satisfy the following properties
\begin{eqnarray}
\eta^2=0  \\
(\eta^\dagger)^2=0  \\
\left\lbrace \eta, \eta^\dagger \right\rbrace= 2 	I
\label{eq:anticomm}
\end{eqnarray}
In addition, we require the $\eta$ matrices to be symmetric. We find that the minimum dimension for the matrices $\eta$ that are symmetric and satisfy the relations above is four. Moreover, these matrices were found as pairs of two, one being the conjugate of the other. Therefore, the matrices $\eta$ are symmetric 4$\times$4 nilpotent matrices and the only eigenvalue of these matrices is 0. Furthermore, the trace and determinant of these matrices are also zero. The wave function $\psi$ in equation (\ref{mse-eq}) is therefore a column matrix with four components. There are several possible representations of the matrix $\eta$. For instance, two possible representations of this matrix are as follows
\begin{eqnarray}
\eta &=& 
\frac{1}{\sqrt{2}}
\left(
\begin{array}{cccc}
 0 & -i & 0 & -1 \\
 -i & 0 & 1 & 0 \\
 0 & 1 & 0 & -i \\
 -1 & 0 & -i & 0
\end{array}
\right)
=
\frac{-i}{\sqrt{2}}\left(
\begin{array}{cc}
  \sigma _1 &  \sigma _2 \\
 -\sigma _2 &  \sigma _1
\end{array}
\right)
\label{eq:matrix-rep1}
\end{eqnarray}
and
\begin{eqnarray}
\eta &=& 
\frac{i}{\sqrt{2}}\left(
\begin{array}{cc}
 0 &  I+\sigma _2 \\
 I-\sigma _2 &  0
\end{array}
\right)
\label{eq:matrix-rep2}
\end{eqnarray}
where $\sigma_i$ are the Pauli matrices. The matrix $\eta$ is symmetric ($\eta^T = \eta$) but not Hermitian, therefore, $\eta^\dagger=\eta^* \neq \eta$. For our analysis herein, we choose the representation in equation (\ref{eq:matrix-rep1}). The choice of this representation is primarily because the eigenstates corresponding to this representation are more relevant for the analysis in the following sections.  

\subsubsection*{Probability Current} 

Multiplying equation (\ref{mse-eq}) with $ \psi^\dagger (\eta+\eta^\dagger)$ from the left we obtain the following equation
\begin{eqnarray}
-i \psi^\dagger(  \eta+ \eta^\dagger) \partial_z  \psi = (i  \psi^\dagger \eta^\dagger  \eta \partial_t ) \psi + m \psi^\dagger \eta \eta^\dagger \psi
\label{eq:prob-1}
\end{eqnarray}
Similarly, taking the complex conjugate of equation (\ref{mse-eq}) and multiplying by $(\eta+ \eta^\dagger) \psi$ on the right we obtain the following equation 
\begin{eqnarray}
i \partial_z \psi^\dagger (\eta+ \eta^\dagger) \psi = (-i  \partial_t \psi^\dagger \eta^\dagger  \eta  ) \psi + m \psi^\dagger \eta \eta^\dagger \psi
\label{eq:prob-2}
\end{eqnarray}
Subtracting (\ref{eq:prob-1}) from (\ref{eq:prob-2}) we obtain the continuity equation
\begin{eqnarray}
i \partial_z [(\psi^\dagger ( \eta+\eta^{\dagger} ) \psi)] = -i  \partial_t [\psi^\dagger \eta^\dagger  \eta  \psi ]
\label{eq:prob-3}
\end{eqnarray}
with the probability and current densities given as
\begin{eqnarray}
J &=& \psi^\dagger ( \eta+\eta^{\dagger} ) \psi \\
\rho &=&  \psi^\dagger \eta^\dagger \eta \psi
\end{eqnarray}
The probability and current densities are therefore Hermitian. For the representation of $\eta$ in (\ref{eq:matrix-rep1}), $\eta+\eta^\dagger=-i\sqrt{2}\gamma_2$ and $\eta^\dagger\eta=I+i\gamma_3$, where $\gamma_i$ are the gamma matrices.

\subsection{Plane Wave Solutions}

We first consider motion along a single ($z$) direction and seek plane wave solutions of equation (\ref{mse-eq}) of the form
\begin{eqnarray}
\psi=u(p) e^{-i p.x}=u(p) e^{-i (E t - p_z z )}
\end{eqnarray}
where, $E= p_z^2/2m$. Therefore, in momentum space equation (\ref{mse-eq}) becomes
\begin{eqnarray}
p_z \ \psi = (E \eta  +m \eta^\dagger ) \ \psi 
\label{eq:momentum-space}
\end{eqnarray}
The above equation is an eigenvalue equation with the momentum operator given as, $\hat{P}=E \eta  + \eta^\dagger m$, which is not Hermitian. However,  the eigenvalues of this operator are real. The eigenvectors of the matrix $E \eta  + m \eta^\dagger$ consists of two states with eigenvalues of momentum $+ \sqrt{2Em}=p_z$ and two other with states $-\sqrt{2Em}=-p_z$. This is consistent with the energy momentum relationship of particles in quantum mechanics, $E=p_z^2/2m$. Following are the eigenstates of the matrix $E \eta  + \eta^\dagger m$,
\begin{eqnarray}
u^{(1)}=\left(
\begin{array}{c}
 1 \\
 0 \\
 i \alpha (E-m) \\
 - \sqrt{2} \alpha  p_z 
\end{array}
\right) , \ \ \ \ 
u^{(2)}=\left(
\begin{array}{c}
 0 \\
 1 \\
    \sqrt{2}\alpha p_z \\
 -i \alpha (E-m)
\end{array}
\right) \\
\label{eq:eigenstates-12}
u^{(3)}=\left(
\begin{array}{c}
 1 \\
 0 \\
 i \alpha (E-m) \\
  \sqrt{2}\alpha  p_z
\end{array}
\right), \ \ \ \
u^{(4)}=\left(
\begin{array}{c}
 0 \\
 1 \\
 -   \sqrt{2}\alpha p_z \\
 -i \alpha (E-m)
\end{array}
\right),
\label{eq:eigenstates-34}
\end{eqnarray}
where $\alpha=1/(E+m)$ and $p_z=\sqrt{2 E m}$. The eigenstates $u^{(1)}$ corresponds to a spin up particle with positive momentum and $u^{(2)}$ corresponds to spin down particle with the same momentum. The eigenstates $u^{(3)}$ and $u^{(4)}$ correspond to particles with negative momentum, i.e., moving in the negative $z$ direction. The eigenstates are related as $u^{(1)}(p_z)=u^{(3)}(-p_z)$ and $u^{(2)}(p_z)=u^{(4)}(-p_z)$. We will employ these eigenstates in the following section when we study the scattering problem. The normalization condition for the eigenstates are
\begin{eqnarray}
u^{{(1)} \dagger} u^{(1)}=2 \\
u^{{(1)} \dagger} u^{(2)}=0
\end{eqnarray}
and similarly for $u^{(3)}$ and $u^{(4)}$. Since the energy momentum relationship of these states does not follow the  relativistic expression, it is not relevant to discuss how these states transform under Lorentz transformations. 

\section{Potential Step problem}\label{step}

\begin{figure}
\centering
\includegraphics[scale=.5]{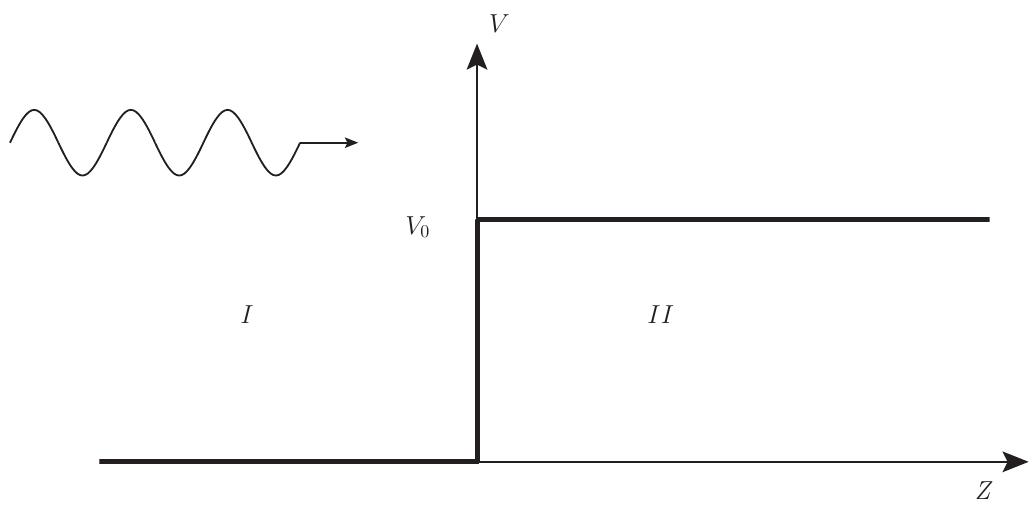}
\caption{Particle incident on a step potential.}
\label{fig:step}
\end{figure}

The potential step problem is one of the foundational examples in quantum mechanics (Figure \ref{fig:step}). As described earlier, solving this problem with the Dirac equation leads to the Klein paradox. In this section we analyze this problem for electron scattering from a potential step in 1 dimension using the solutions presented in previous sections. We will show that the resulting transmission and reflection coefficients for spin up and down particles are related to the ones in quantum mechanics. For regions $I$ and $II$, equation (\ref{eq:momentum-space}) is given by
\begin{align}
p_1 \ \psi =& (E \eta  +m \eta^\dagger ) \ \psi  & (I) \label{eq:inc-wf} \\
p_2 \ \psi =& ((E-V_0) \eta  +m \eta^\dagger ) \ \psi & (II)  \label{eq:ref-wf}
\end{align}

\subsection{Case I: $E>V_0$}

We first consider the case of a spin up electron incident on a potential step with energy $E>V_0$. The incident and reflected electron waves in region $I$ are given by
\begin{eqnarray}
\psi_I=A \left(
\begin{array}{c}
 1 \\
 0 \\
 i \alpha (E-m) \\
 - \sqrt{2} \alpha  p_1 
\end{array}
\right) e^{i p_1 z}
\end{eqnarray}

\begin{eqnarray}
\psi_I^r=B\left(
\begin{array}{c}
 1 \\
 0 \\
 i \alpha (E-m) \\
  \sqrt{2} \alpha  p_1 
\end{array}
\right)  e^{-i p_1 z} +
B' \left(
\begin{array}{c}
 0 \\
 1 \\
  -  \sqrt{2}\alpha p_1 \\
 -i \alpha (E-m)
\end{array}
\right)  e^{-i p_1 z}
\end{eqnarray}

\begin{figure}
\vspace*{-1.2cm}
\centering
\includegraphics[scale=.4]{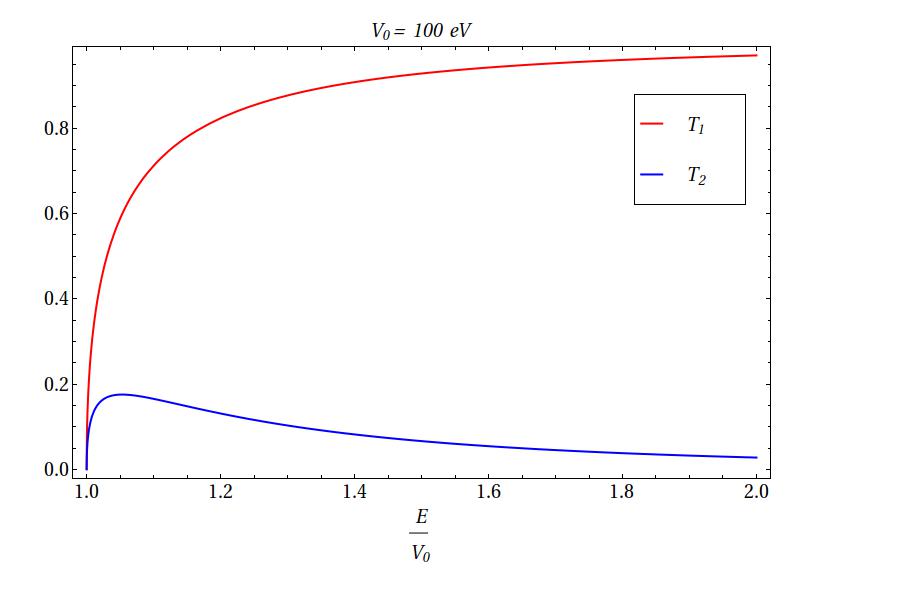}
\includegraphics[scale=.4]{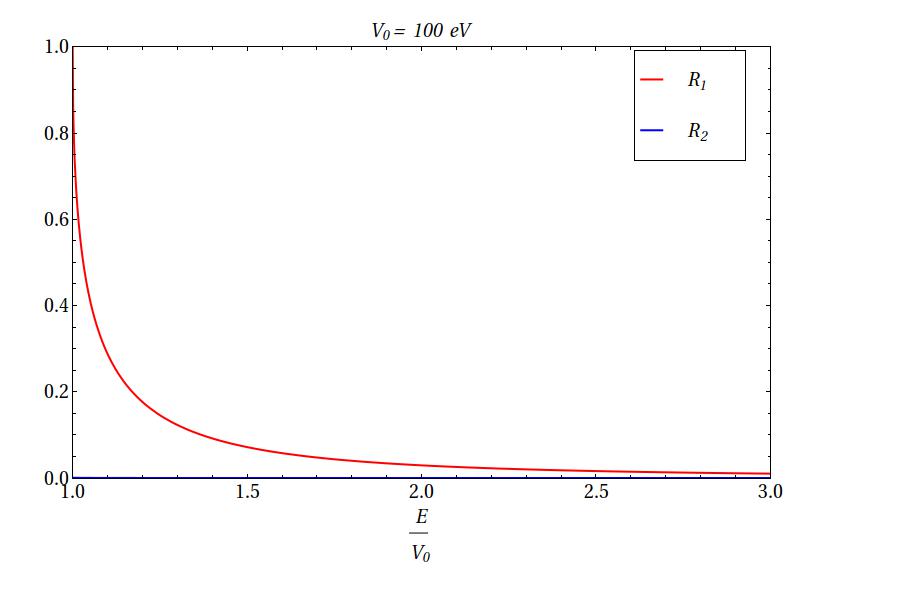}
\caption{The plot shows the transmission and reflection coefficients for spin up and down electrons given in equations (\ref{eq:t1}), (\ref{eq:t2}), (\ref{eq:r1}) and (\ref{eq:r2}). The red lines show the coefficients of spin up electron and the blue lines correspond to spin down electron. The value of the potential step is chosen to be $V_0=$100 eV. For $E \simeq V_0$ there is no transmission and the electron is completely reflected with its spin up. For $E$ slightly greater than $V_0$, the transmission of spin up and down electron occurs with nearly equal probabilities. For $E/V_0 \gtrsim 1.05$ the transmitted electron is dominantly spin up. The lower panel shows the reflection coefficients. We can see that there is  a very small probability that the reflected electron has its spin flipped. The reflection coefficient for spin down is small but non-zero. As the energy increases the barrier appears more and more transparent to the electron and the transmission coefficient of the spin up electron becomes large. For $E \simeq 3V_0$ the barrier is nearly transparent to the electron.}
\label{fig:transmission-100ev}
\end{figure}

\begin{figure}
\centering
{\includegraphics[scale=.4]{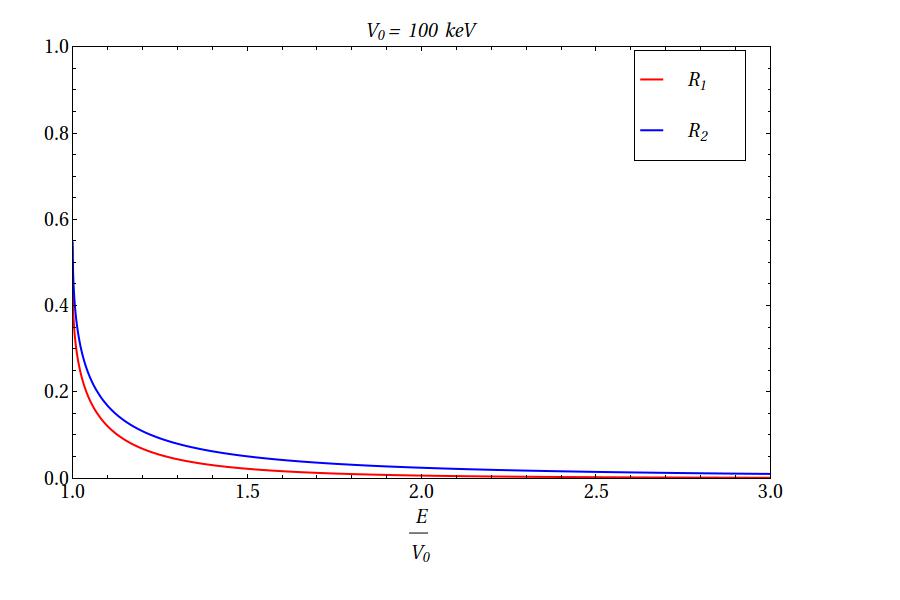}}
{\includegraphics[scale=.4]{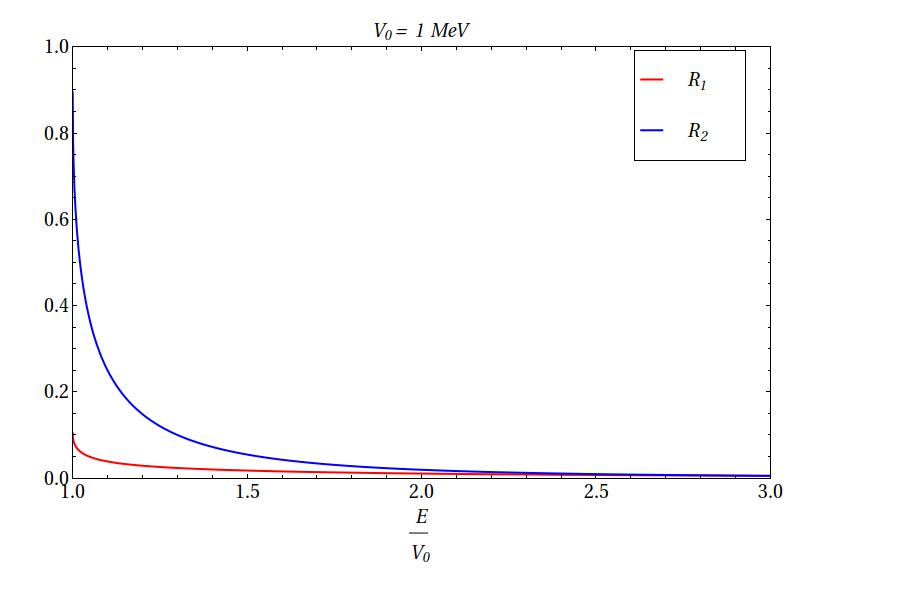}}
\caption{Plot shows the reflection coefficients (\ref{eq:r1}) and (\ref{eq:r2}) for $V_0=$ 100 keV and $V_0=$1 MeV. The red lines correspond to spin up electrons whereas the blue lines are for spin down electron. The plot of the transmission coefficients for these cases are similar to the one shown in Figure \ref{fig:transmission-100ev}. We can observe that the  reflection coefficient changes notably as the height of the barrier is increased. For $V_0=$100 keV, the probability of the spin up electron being reflected as a spin up or down electron is nearly equal. For $V_0=$1 MeV,  the electron is dominantly reflected with its spin flipped.}
\label{fig:reflection-1mev}
\end{figure}

Similarly for region $II$ we have 
\begin{eqnarray}
\psi_{II}=C\left(
\begin{array}{c}
 1 \\
 0 \\
 i \beta (E-V_0-m) \\
 - \sqrt{2} \beta  p_2 
\end{array}
\right)  e^{i p_2 z} +
C' \left(
\begin{array}{c}
 0 \\
 1 \\
    \sqrt{2}\beta p_2 \\
 -i \beta (E-V_0-m)
\end{array}
\right)  e^{i p_2 z}
\end{eqnarray}
where $\alpha=1/(E+m)$, $\beta=1/(E-V_0+m)$, $p_1=\sqrt{2Em}$ and $p_2=\sqrt{2(E-V_0)m}$. At $z=0$ the continuity of the wave function implies
\begin{eqnarray}
\psi_I(z=0)+\psi_{I}^r(z=0)=\psi_{II}(z=0)
\label{eq:cont-cond}
\end{eqnarray}
From this condition the coefficients are found to be
\begin{eqnarray}
 \frac{B}{A} &=& \frac{(-{E}+m) {V_0}}{({E}+m) \left(2 {E}+2 \sqrt{{E} ({E}-{V_0})}-{V_0}\right)} \\
B'&=& C'\\[5mm]
\frac{C}{A} &=& \frac{-2 {E}^2-2 m \sqrt{{E} ({E}-{V_0})}+2 {E} \left(m+\sqrt{{E} ({E}-{V_0})}+{V_0}\right)}{({E}+m) {V_0}}\\
\frac{C'}{A} &=& \frac{2 i \sqrt{{E} m} {V_0}}{({E}+m) \left(2 {E}+2 \sqrt{{E} ({E}-{V_0})}-{V_0}\right)}
%
\end{eqnarray}
The transmission and reflection coefficients $T_1, \ T_2, \ R_1$ and $R_2$ are given by
\begin{eqnarray}
T_1 &=& \frac{(E+m)(E-{V_0})^{1/2}}{(E-{V_0}+m)E^{1/2}}\left| \frac{C}{A} \right|^2 , \ \
T_2 = \frac{(E+m)(E-{V_0})^{1/2}}{(E-{V_0}+m)E^{1/2}}\left| \frac{C'}{A} \right|^2 \\[1cm]
R_1 &=& \left| \frac{B}{A} \right|^2 , \ \ 
R_2 = \left| \frac{B'}{A} \right|^2 
\end{eqnarray}
The complete expression for these coefficients are presented in the appendix in equations (\ref{eq:t1}), (\ref{eq:t2}), (\ref{eq:r1}) and (\ref{eq:r2}). The sum of these coefficient is always equal to 1, i.e.,
\begin{eqnarray}
 (T_1+T_2)+ (R_1+R_2) =1
\end{eqnarray}
The quantum mechanical transmission and reflection coefficients are related to these coefficients as 
\begin{eqnarray}
T_{QM} &=& T_1+T_2 =\frac{ 4 E^{1/2}\sqrt{E-V_0}}{\left(\sqrt{E}+\sqrt{E-V_0}\right)^2} \\
R_{QM} &=& R_1+R_2 =\left(\frac{\sqrt{E}-\sqrt{E-V_0}}{\sqrt{E}+\sqrt{E-V_0}}\right)^2 
\end{eqnarray}
We can see that although the reflection and transmission coefficients ($T_1$, $T_2$, $R_1$ and $R_2$) depend on the mass of the incident particle, the sum, which is the quantum mechanical result, is independent of it. The current densities for the incident, reflected and transmitted spin up and down particle are given by
\begin{eqnarray}
J_I^\uparrow &=& \psi_I^\dagger ( \eta+\eta^{\dagger} ) \psi_I =  \frac{4 |A|^2 p_1 }{E+m} \\
J_r^\uparrow &=& \psi_{r \uparrow}^\dagger ( \eta+\eta^{\dagger} ) \psi_{r \uparrow} =   -\frac{4 {p_1} |B|^2 }{E+m}\\
J_r^\downarrow &=& \psi_{r \downarrow}^\dagger ( \eta+\eta^{\dagger} ) \psi_{r \downarrow} =   -\frac{4 {p_1} |B'|^2 }{E+m}\\
J_T^\uparrow &=& \psi_{T \uparrow}^\dagger ( \eta+\eta^{\dagger} ) \psi_{T \uparrow} =   \frac{4 {p_2} |C|^2 }{E+m-{V_0}} \\
J_T^\downarrow &=& \psi_{T \downarrow}^\dagger ( \eta+\eta^{\dagger} ) \psi_{T \downarrow} =   \frac{4 {p_2} |{C'}|^2}{E+m-{V_0}} 
\end{eqnarray}
The conservation of probability density implies
\begin{eqnarray}
\frac{|J_r^\uparrow|}{|J_I^\uparrow|}+\frac{|J_T^\uparrow|}{|J_I^\uparrow|}+
\frac{|J_r^\downarrow|}{|J_I^\uparrow|}+\frac{|J_T^\downarrow|}{|J_I^\uparrow|}
=1
\end{eqnarray}
\begin{eqnarray}
\frac{|B|^2}{|A|^2}+\frac{(E+m)(E-{V_0})^{1/2}}{(E-{V_0}+m)E^{1/2}} \frac{|C|^2}{|A|^2}+
\frac{|B'|^2}{|A|^2}+\frac{(E+m)(E-{V_0})^{1/2}}{(E-{V_0}+m)E^{1/2}} \frac{|C'|^2}{|A|^2}=1
\end{eqnarray}
which agrees with the calculations of the transmission and reflection coefficient. Figure \ref{fig:transmission-100ev} shows the plot of the transmission and reflection coefficients. We can see from this Figure that when the energy of the incident electron is close to the height of the barrier the electron is completely reflected. As the energy of the electron increases relative to the barrier the reflection coefficient falls sharply and the electron is mostly transmitted as spin up. There is however a small probability that the transmitted electron flips its spin as well. This can be seen from the blue line in the upper panel of Figure \ref{fig:transmission-100ev}. We show the reflection coefficients for $V_0= 100 \ keV$ and $1 \ MeV$ in Figure \ref{fig:reflection-1mev}. We can see that the spin of the reflected electron depends on the height of the potential barrier. For $V_0 < m_e$  the reflected electron is dominantly spin up (Figure \ref{fig:transmission-100ev}) whereas for $V_0=1 \ MeV  > m_e$, the electron is mostly reflected with spin down. The upper panel of Figure \ref{fig:reflection-1mev} shows an intermediate situation when the probabilities of reflection of electron as spin up and down are comparable.

\subsection{Case II: $E<V_0$}

\begin{figure}
\centering
\includegraphics[scale=.35]{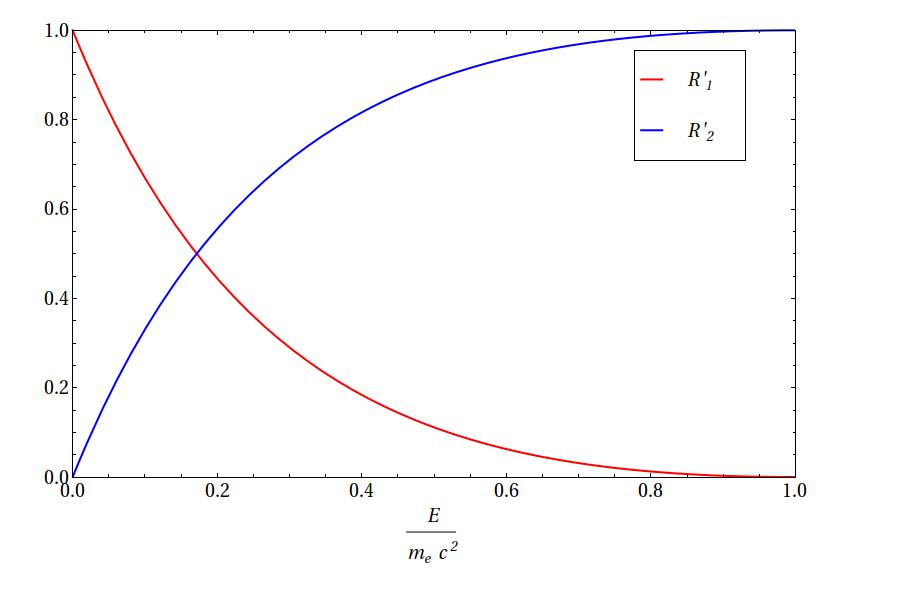}
\caption{Figure shows the plot of the reflection coefficients for the case when energy of the incident electron is less than the height of the barrier, $E<V_0$. The coefficients are given in equations (\ref{eq:reflection-eltv0-up}) and  (\ref{eq:reflection-eltv0-dn}). For very small values of energy the incident spin up electron is reflected without a change in its spin. As the energy of the incident electron increases the probability that it flips its spin upon reflection increases considerably.}
\label{fig:relection-eltv0}
\end{figure}

We next analyze the case when the energy of the incident electron is less than the height of the barrier. The equations for incident and reflected electron in region $I$  remain the same as equations (\ref{eq:inc-wf}) and (\ref{eq:ref-wf}). For the transmitted electron in region $II$, however, the equation is now given by
\begin{align}
i p'_2 \ \psi =& [-(V_0-E) i\eta  +m i\eta^\dagger ] \ \psi & (II)  
\end{align}
The operator $\hat{P}=-(V_0-E) i\eta  +m i\eta^\dagger$ has eigenvalues $\pm i\sqrt{2(V_0-E)m}=\pm i p'_2$. We choose the eigenvalue $i p'_2$ which imply a decaying wave ($\psi=u(p) e^{-i p.x}=u(p) e^{-i E t } e^{- p_z z}$) within the barrier. We choose the eigenvectors corresponding to eigenvalue $i p'_2$ for spin up and down electrons in region $II$. The wave function in region $II$ is therefore given by
\begin{eqnarray}
\psi'_{II}=D\left(
\begin{array}{c}
 1 \\
 0 \\
 i \rho (V_0-E+m) \\
  \sqrt{2} i \rho \  p'_2 
\end{array}
\right)  e^{- p'_2 z} +
D' \left(
\begin{array}{c}
 0 \\
 1 \\
    -\sqrt{2} i \rho \ p'_2 \\
 -i \rho (V_0-E+m)
\end{array}
\right)  e^{- p'_2 z}
\end{eqnarray}
where $\rho=1/(V_0-E-m)$ and $p'_2=\sqrt{2m(E-V_0)}$. From the continuity condition (\ref{eq:cont-cond}) we find the coefficients for this case as well. These are presented in equations (\ref{eq:t1b}), (\ref{eq:t2b}) and (\ref{eq:r1b}) and (\ref{eq:r2b})  of the Appendix.
The reflection coefficients $R'_1=|B/A|^2$ and $R'_2=|B'/A|^2$  coefficients are given by
\begin{eqnarray}
R'_1=\frac{(E-m)^2}{(E+m)^2}
\label{eq:reflection-eltv0-up}\\
R'_2=\frac{4 E m}{(E+m)^2}
\label{eq:reflection-eltv0-dn}
\end{eqnarray}
The sum of these coefficients is always equal to one. The current densities  for the transmitted wave ($J_T^\uparrow$ and $J_T^\downarrow$) is zero in this case and therefore there is essentially no wave transmitted in the barrier. Moreover, the reflection coefficients are independent of the height of the barrier for $E<V_0$. The probability of the reflection of spin up and down electrons is shown Figure \ref{fig:relection-eltv0}. We can see that for very small energies the incident spin up electron is reflected with the same spin. As the energy of the electron increases, the probability that it is reflected with the spin flipped increases considerably.

\section{Extending to 3 Dimensions}\label{3d}

In the previous sections we saw how solutions of equation (\ref{mse-eq}) leads to the transmission and reflection coefficients of the spin up and down electron scattering for the case of an electron scattering off a potential step. The analysis was performed in 1 dimension. For 3 dimensions the analysis is more subtle and here we discuss a version of equation (\ref{mse-eq}) for 3 dimensions.

Following is the 3 dimensional version of equation (\ref{mse-eq}) 
\begin{eqnarray}
-i \mu_i \partial_i \psi = (i  \eta \partial_t  + \eta^\dagger m) \psi 
\label{mse-eq-3d}
\end{eqnarray}
Here,  $\mu_i \partial_i=\mu_1 \partial_1+\mu_2 \partial_2+\mu_2 \partial_2$ ($i=1,2,3$) and the matrices $\mu_i$ are given by
\begin{eqnarray}
\mu_1 &=&
i \gamma_1 \gamma_2 
=
\left(
\begin{array}{cc}
  \sigma _3 &  0 \\
 0 &  \sigma _3
\end{array}
\right)
\label{eq:matrix-mu1}
\\
\mu_2 &=&
 \gamma_0 \gamma_2 
=
\left(
\begin{array}{cc}
  0 &  \sigma _2 \\
 \sigma _2 & 0
\end{array}
\right)
\label{eq:matrix-mu2}
\\
\mu_3 &=&
 \gamma_2 \gamma_5 
=
\left(
\begin{array}{cc}
  \sigma _2  & 0 \\
 0 & -\sigma _2 
\end{array}
\right)
\label{eq:matrix-mu3}
\end{eqnarray}
These matrices are Hermitian and obey the following anticommutation relation
\begin{eqnarray}
\left\lbrace \mu_i, \mu_j \right\rbrace= 2 \delta_{ij} 	I
\label{eq:anticomm-3d}
\end{eqnarray}
Squaring the 3D equation and using (\ref{eq:anticomm-3d}) along with the properties of the $\eta$ matrices then leads to the Schrodinger equation in 3D.  The continuity equation can be obtained using a similar procedure employed in section \ref{equation}. For this case we first multiply $ \psi^\dagger \gamma_3 (\eta+\eta^\dagger)$ from the left of (\ref{mse-eq-3d}) and then multiply $(\eta+ \eta^\dagger) \gamma_3 \psi$ to the right with the complex conjugated equation.  Next, using the relations $[\mu_i,\gamma_3]=0$, $\{ \mu_i,  \eta+\eta^\dagger \}=0 $, $[\eta\eta^\dagger,\gamma_3]=0$ and $\{ \eta+\eta^\dagger,\gamma_3\}=0 $ we obtain the following continuity equation
\begin{eqnarray}
 \partial_i [\psi^\dagger \Sigma_{i} \psi] = -  \partial_t [\psi^\dagger  \Gamma \psi ]
\label{eq:prob-3}
\end{eqnarray}
where $\Sigma_{i}=i \mu_i ( \eta+\eta^{\dagger} )\gamma_3$ and $\Gamma=i \eta^\dagger  \eta \gamma_3$ are Hermitian matrices. The probability and current densities for this case are therefore given by
\begin{eqnarray}
J_{i} &=& \psi^\dagger \Sigma_{i} \psi\\
\rho &=&  \psi^\dagger \Gamma \psi
\end{eqnarray}
Next, as before, we can seek plane wave solutions and obtain the 3D equation in momentum space as
\begin{eqnarray}
 \mu_i p_i \psi = ( \eta E  + \eta^\dagger m) \psi 
\label{mse-eq-2d}
\end{eqnarray}
which is the relevant equation for studying 3D scattering problems. The application of this equation to such problems in 3D is more involved and we leave it for future investigation.

Lastly, we comment on the Pauli equation which describes the interaction of a spin 1/2 particle with an external electromagnetic field. It is obtained in the non-relativistic limit of the Dirac equation by assuming the presence of an electromagnetic field. Furthermore, it correctly predicts the spin of the particle and the gyromagnetic ratio. The Schrodinger equation, however, does not predict the spin of the particle. The fundamental first order equation introduced in this article is non-relativistic and it also does not  predict the spin of the particle. Moreover, the Pauli equation cannot be obtained from this equation. However, this equation does include the spin of the particle in scattering problems which cannot be implemented using the Schrodinger equation.

\section{Conclusion}\label{conclude}

We proposed that the Schrodinger equation can be derived from a more fundamental first order equation in one and three dimensions, given in (\ref{mse-eq}) and (\ref{mse-eq-3d}). Nilpotent symmetric matrices that obey  anticommutation relation (\ref{eq:anticomm}) were introduced for this purpose. We further constructed the solutions of this equation in 1D and employed these to solve the problem of a spin up electron scattering from a potential step. We showed that the quantum mechanical transmission and reflection coefficients are obtained as the sum of those for the spin up and down electron. Experimental tests are required to examine the predictions of this analysis. Moreover, further investigation is also needed by applying this equation to other similar problems in quantum mechanics.

\section*{Appendix A} 
\setcounter{equation}{0}  

\renewcommand{\theequation}{A-\arabic{equation}}

\subsection*{Transmission and Reflection Coefficients for $E>V_0$}
\begin{eqnarray}
T_1 &=& \frac{(E+m)(E-{V_0})^{1/2}}{(E-{V_0}+m)E^{1/2}}\left| \frac{C}{A} \right|^2  \nonumber \\
 &=& \frac{4 \sqrt{1-\frac{{V_0}}{E}} \left(E^2+m \sqrt{E (E-{V_0})}-E \left(m+\sqrt{E (E-{V_0})}+{V_0}\right)\right)^2}{(E+m) (E+m-{V_0}) {V_0}^2} \label{eq:t1}\\[1cm]
T_2 &=& \frac{(E+m)(E-{V_0})^{1/2}}{(E-{V_0}+m)E^{1/2}}\left| \frac{C'}{A} \right|^2 \nonumber \\[1cm]
 &=& \frac{4 m \sqrt{E (E-{V_0})} {V_0}^2}{(E+m) (E+m-{V_0}) \left(-2 E-2 \sqrt{E (E-{V_0})}+{V_0}\right)^2} \label{eq:t2}
\end{eqnarray}
\begin{eqnarray}
R_1 &=& \left| \frac{B}{A} \right|^2 \nonumber \label{eq:r1}  \\ 
&=& \frac{(E-m)^2 {V_0}^2}{(E+m)^2 \left(-2 E-2 \sqrt{E (E-{V_0})}+{V_0}\right)^2} \\
R_2 &=& \left| \frac{B'}{A} \right|^2  \nonumber \label{eq:r2}  \\ 
&=& \frac{4 E m {V_0}^2}{(E+m)^2 \left(-2 E-2 \sqrt{E (E-{V_0})}+{V_0}\right)^2}
\end{eqnarray}
\subsection*{Coefficients for $E<V_0$}

\begin{eqnarray}
\frac{B}{A} &=& \frac{(-E+m) {V_0}}{(E+m) \left(2 E-{V_0}+2 i \sqrt{E (-E+{V_0})}\right)} \label{eq:t1b}\\
\frac{B'}{A} &=& \frac{2 \sqrt{E m} {V_0}}{(E+m) \left(-2 i E+i {V_0}+2 \sqrt{E (-E+{V_0})}\right)}  \label{eq:t2b} \\
\frac{D}{A} &=& \frac{2 \left(E^2+E (m-{V_0})+i \left(m \sqrt{E (-E+{V_0})}+\sqrt{E^3 (-E+{V_0})}\right)\right)}{(E+m) \left(2 E-{V_0}+2 i \sqrt{E (-E+{V_0})}\right)}  \label{eq:r1b}\\
\frac{D'}{A} &=&\frac{2 \sqrt{E m} {V_0}}{(E+m) \left(-2 i E+i {V_0}+2 \sqrt{E (-E+{V_0})}\right)}  \label{eq:r2b}
\end{eqnarray}

\end{document}